\documentclass[12pt]{article}

\usepackage{amsmath}
\usepackage{amssymb}
\usepackage{amsthm}

\theoremstyle{plain}
\newtheorem{theorem}{Theorem}
\newtheorem{lemma}{Lemma}



\usepackage{babel}

\begin{document}
\title{\textbf{Power laws and logarithmic oscillations in diffusion processes on infinite countable ultrametric spaces }}
\author{A.\,Kh.~Bikulov \\
 \textit{Institute of Chemical Physics, } \\
 \textit{Kosygin street 4, 117734 Moscow, Russia} \\
 e-mail:\:\texttt{beecul@mail.ru} \\
 and \\
 A.\,P.~Zubarev \\
 \textit{ Physics Department, Samara University, } \\
 \textit{ Moskovskoe shosse 34, 443123, Samara, Russia} \\
 \textit{Natural Science Department, } \\
 \textit{Samara State University of Railway Transport,} \\
 \textit{Perviy Bezimyaniy pereulok 18, 443066, Samara, Russia} \\
 e-mail:\:\texttt{apzubarev@mail.ru} }
\maketitle
\begin{abstract}
It is shown that if the initial condition of the Cauchy problem
for the diffusion equation on a general infinite countable
ultrametric space is spherically symmetric with respect to some
point, then this problem has an exact analytical solution. A
general solution of this problem is presented for pure ultrametric
diffusion, as well as for ultrametric diffusion with a reaction
sink concentrated at the center of spherical symmetry. Conditions
on the ultrametric and the distribution of the number of states in
ultrametric spheres are found that lead at large times to the
asymptotic behavior of the solutions obtained in the form of a
power law modulated by a bounded function that is log-periodic
under some additional conditions.

\vspace{5mm}
\end{abstract}

\section{Introduction}

\label{sec_1}

A characteristic feature of models based on $p$-adic diffusion
described by the Vladimirov equation \cite{VVZ} is the presence of
power relaxation laws in these models. In the recent paper
\cite{BZ_2023}, the authors, motivated by \cite{Nonnenmacher,MKJ},
have shown that the solution of the Cauchy problem with the
initial condition on a compact set for pure $p$-adic diffusion and
for the case of $p$-adic diffusion with a reaction sink is
described at large times by a power law modulated by a
logarithmically oscillating function, i.e., by a relation of the
form

\begin{equation}
t^{-\beta}f\left(t\right),\label{log_per}
\end{equation}
where $\beta>0$ and $f\left(t\right)$ is a bounded log-periodic
function: $f\left(\kappa t\right)=f\left(t\right)$ for some
$\kappa>1$. It is important to note that a similar behavior of the
dynamical variables for a number of physical, chemical,
biological, and socio-economic systems has been known since the
early 1990s (see, for example,
\cite{Blatz,Anifrani,Sahimi,Johansen_1,Feigenbaum,Gluzman,Drozdz,Sornette})
and was associated with the discrete scale invariance of such
systems \cite{Saleur_1,Saleur_2}. As pointed out in
\cite{BZ_2023}, a similar behavior of relaxation curves is
characteristic of models of $p$-adic diffusion and indicates the
discrete scale invariance of the equations of these models. The
present paper is a further development of \cite{BZ_2023} and is
devoted to the analysis of conditions under which power laws and
lows of the form (\ref{log_per}) arise in ultrametric diffusion
processes on general infinite countable ultrametric spaces, which
generally speaking are not subsets of the field of $p$-adic
numbers.

It is well known (see, for example, the reviews
\cite{ALL,ALL_1,ALL_2,Koz}) that the process of $p$-adic diffusion
underlies the ultrametric approach to the description of
conformational dynamics of protein in the native state. In this
approach, the set of conformational states of protein $\left\{
C\right\} $ is identified with the set of its quasi-equilibrium
states. Here each quasi-equilibrium state is associated with a
local minimum of the potential energy of protein as a function of
all its generalized coordinates. It can be shown (see, for
example, \cite{Koz,Koz_2010,BZ_2021}) that, on the set of local
minima, one can naturally introduce a distance function (a
metric), which is an ultrametric; thus, the set of conformational
states of protein $\mathcal{U}=\left\{ C\right\} $ is an
ultrametric space. Since the mean sojourn time of protein in
quasi-equilibrium states is much larger than the transition times
between these states \cite{BZ_2021}, the protein dynamics in the
space of such states can be described by a discrete-time Markov
random process in which only the sojourn times of protein in the
states are taken into account. In this description, the
distribution function $f_{C}\left(t\right)$ of protein over the
set $\mathcal{U}$ satisfies the Kolmogorov--Feller equation
(master equation) \cite{Gardiner}
\begin{equation}
\dfrac{df_{C}\left(t\right)}{dt}=\sum_{C^{\prime}\in\mathcal{U}}\left(P_{CC^{\prime}}f_{C^{\prime}}\left(t\right)-P_{C^{\prime}C}f_{C}\left(t\right)\right),\label{KF_Conf}
\end{equation}
where $P_{CC^{\prime}}$ is the probability of transition in unit
time from state $C^{\prime}$ to state $C$. To describe experiments
on the kinetics of ligand binding to myoglobin and spectral
diffusion experiments (see
\cite{BZ_2021,AB_2008,ABKO_2002,ABZ_2014,ABO_2004}), it suffices
to use an approximation in which all quasi-equilibrium states have
the same configuration volume and the same free energy. Moreover,
within the approximation used, we assume that the energy barrier
$E\left(C,C^{\prime}\right)$ between states $C$ and $C^{\prime}$
is a linear function of the ultrametric distance
$d\left(C,C^{\prime}\right)$ between these states, i.e.,
$E\left(C,C^{\prime}\right)=E_{0}d\left(C,C^{\prime}\right),$
where the parameter $E_{0}>0$ defines the energy scale. Within
these approximations, the probability of transition from state
$C^{\prime}$ to state $C$ in unit time has the form
\begin{equation}
P_{CC^{\prime}}=\dfrac{1}{\tau}\dfrac{1}{N\left(E\left(C,C^{\prime}\right)\right)}\exp\left(-\dfrac{E_{0}d\left(C,C^{\prime}\right)}{kT}\right),\label{P_CC}
\end{equation}
where $N\left(E\left(C,C^{\prime}\right)\right)\equiv
N\left(d\left(C,C^{\prime}\right)\right)$ is the number of states
in the subset of all states that contains $C$ and $C^{\prime}$ as
well as all states separated from each other by energy barriers
not higher than $E\left(C,C^{\prime}\right)$, $\tau$ is a
dimensional factor that determines the time scale, $T$ is
temperature, and $k$ is the Boltzmann constant. Equation
(\ref{KF_Conf}) with the unit-time transition probability
(\ref{P_CC}) is called an ultrametric diffusion equation. Formula
(\ref{P_CC}) has the following simple physical interpretation. The
probability $P_{CC^{\prime}}$ should be proportional to the
Boltzmann factor $\exp\left(-\tfrac{E\left(C,C^{\prime}\right)}{kT}\right)$, which
determines the probability that the system receives energy equal
to the energy barrier $E\left(C,C^{\prime}\right)$. In this case,
if the system has energy less than $E\left(C,C^{\prime}\right)$,
then it can in principle pass to any state separated from the
initial state by an energy barrier not higher than
$E\left(C,C^{\prime}\right)$. Since all quasi-equilibrium states
in this model have the same free energy, the transition
probability to such a state should also contain the factor
$N\left(E\left(C,C^{\prime}\right)\right)$ in the denominator.
Note that, in view of (\ref{P_CC}), the steady-state solution of
equation (\ref{KF_Conf}) is constant and does not depend on the
state $C$ in accordance with the Boltzmann law and the assumptions
of the equality of the configuration volumes and free energies of
all quasi-equilibrium states.

Another approximation in ultrametric models of conformational
dynamics of protein is the assumption of self-similarity of the
ultrametric space of states $\mathcal{U}$. By the self-similarity
of the ultrametric space we mean that each ball in this space can
be decomposed into the same number $p$ of maximal subballs nested
into it. This approximation makes it possible to perform a
$p$-adic parameterization of the space $\mathcal{U}$, i.e., to map
this space into the field of $p$-adic numbers $\mathbb{Q}_{p}$.
The possibility of such a mapping follows from the fact that any
infinite countable self-similar ultrametric space $\mathcal{U}$
with ultrametric $d\left(x,y\right)$, $x,y\in\mathcal{U}$, is
isomorphic for $p>1$ to the set of all balls $\left\{
B_{r}\right\} $ of fixed radius $p^{r}$ from $\mathbb{Q}_{p}$.
Under such a parameterization, the set of conformational states of
the system is given by the set $\left\{ B_{r}\right\} $, and the
distribution function of protein over states is a locally constant
function $f\left(x\right)$ on $\mathbb{Q}_{p}$ with local
constancy radius $p^{r}$. When the ultrametric function
$d\left(x,y\right)=\log\left|x-y\right|_{p}$ is taken in the form
$\left|x-y\right|_{p}>p^{r}$ (here $\left|x\right|_{p}$ is the
$p$-adic norm of $x\in\mathbb{Q}_{p}$), equation (\ref{KF_Conf})
with regard to (\ref{P_CC}) turns into the equation of $p$-adic
diffusion on $\mathbb{Q}_{p}/B_{r}$, which, in turn, turns into
the Vladimirov equation \cite{VVZ} in the limit as
$r\rightarrow-\infty$ under an appropriate choice of the factor
$\tau$.

The self-similarity of the ultrametric space that models the space
of conformational states of protein is the attribute of all
$p$-adic models. This raises the important question of how the
properties of ultrametric diffusion models change if one abandons
the assumption of self-similarity of the ultrametric space, as
well as for other possible realizations of the ultrametric
function. To answer this question, one should consider ultrametric
diffusion models on general ultrametric spaces different from
$\mathbb{Q}_{p}$. In the general case a similar problem was
studied in \cite{Koz,Koz_1,Koz_2,Koz_3}, where the authors
developed general methods for solving pseudodifferential equations
on general ultrametric spaces with the use of the basis of
ultrametric wavelets. Using the results of these studies, in this
paper we show that the solution of the Cauchy problem for equation
(\ref{KF_Conf})--(\ref{P_CC}) on an arbitrary infinite countable
ultrametric space $\mathcal{U}$ can be obtained exactly
analytically if the initial condition of the Cauchy problem is
spherically symmetric with respect to the ultrametric about some
arbitrarily chosen fixed point $x_{0}\in\mathcal{U}$. Moreover, we
show that a similar Cauchy problem for the equation
(\ref{KF_Conf})--(\ref{P_CC}) with an additional term responsible
for the reaction sink concentrated at a point
$x_{0}\in\mathcal{U}$ also admits an exact analytical solution.
Thus, having obtained and analyzed exact analytical solutions of
these problems, we can in principle answer the following question:
what constraints should be imposed on the distribution of points
on ultrametric spheres and on the function of ultrametric distance
between ultrametric spheres with a common center in order that the
relaxation curves in the models of ultrametric diffusion described
by a general equation of the form (\ref{KF_Conf}) exhibit
asymptotic behavior described by power laws modulated by bounded
functions, in particular, by functions of the form
(\ref{log_per}). In the present paper, we make an attempt to solve
this problem.

The paper is organized as follows. In Section \ref{sec_2} we
present a general solution of the Cauchy problem for the
ultrametric diffusion equation on a general infinite countable
ultrametric space when the initial distribution is spherically
symmetric with respect to some point. In Section \ref{sec_3} we
use the results of Section \ref{sec_2} to study the asymptotic
solution of the Cauchy problem with the initial condition
concentrated at a point that is the center of spherical symmetry.
We find conditions that should be imposed on the ultrametric and
the distribution of the number of states on ultrametric spheres in
order that the large-time asymptotic behavior of the solution be
described by a power law modulated by some bounded function. We
also find conditions on the ultrametric and the distribution of
the number of states under which the modulating function is
strictly log-periodic. In Section \ref{sec_4} we present a
solution to the problem of random walk on an infinite countable
ultrametric space with a reaction sink concentrated at an
arbitrary fixed point of the ultrametric space. In Section
\ref{sec_5} we study the asymptotic behavior of the solution
obtained in Section \ref{sec_4} in the case when the support of
the initial condition is concentrated at the sink point and define
conditions that lead to the relaxation law in the form of a power
function multiplied by a function that is either bounded or
strictly log-periodic. The final section is devoted to the
discussion and generalization of the results of the present study
formulated in Theorems \ref{th_3} and \ref{th_4}. In Appendices 1,
2, and 3 we give the proofs of Theorems \ref{th_1}, \ref{th_2},
and \ref{th_4}, respectively.

\section{Problem of spherically symmetric ultrametric random walk }

\label{sec_2}

In this section we describe a method for solving the Cauchy
problem for the ultrametric diffusion equation with a spherically
symmetric initial condition. Let $\mathcal{U}=\left\{ x\right\} $
be an ultrametric space consisting of an infinite countable number
of elements $x$, and let $d\left(x,y\right)$, $x,y\in\mathcal{U}$,
be an ultrametric on $\mathcal{U}$. We will refer to the elements
of $\mathcal{U}$ as points of $\mathcal{U}$. We consider the
Cauchy problem on $\mathcal{U}$ for an equation of the form
\begin{equation}
\dfrac{df\left(x,t\right)}{dt}=\left(\hat{A}f\right)\left(x,t\right)\label{Re_Di_Eq}
\end{equation}
with the initial condition
\begin{equation}
f\left(x,0\right)=f_{0}\left(x\right),\label{Re_Di_Nc_gen}
\end{equation}
where the operator $\hat{A}$ is defined as
\begin{equation}
\left(\hat{A}f\right)\left(x\right)=\sum_{y\in\mathcal{U}\setminus x}K\left(d\left(x,y\right)\right)\left(f\left(y\right)-f\left(x\right)\right),\label{Op_A}
\end{equation}
\begin{equation}
K\left(d\left(x,y\right)\right)=\dfrac{1}{N\left(d\left(x,y\right)\right)}\exp\left(-\alpha d\left(x,y\right)\right),\;\alpha=\dfrac{E_{0}}{kT}>0,\label{Ker}
\end{equation}
and $N\left(d\left(x,y\right)\right)$ is the number of points in a
subset of the set $\mathcal{U}$ that also contains the points $x$
and $y$ as well as all points separated from each other by an
ultrametric distance not greater than $d\left(x,y\right)$. It is
obvious that for any ultrametric function $d\left(x,y\right)$ the
operator (\ref{Op_A}) is well defined on functions summable in
absolute value on $\mathcal{U}$.

Since $\mathcal{U}$ is countable, the ultrametric
$d\left(x,y\right)$ can take only a countable number of values.
Suppose that $d_{1}$ is the minimum possible value of the
ultrametric $d\left(x,y\right)$ and all other values
$d_{1},\:d_{2},\:d_{3},\:\ldots$ of the ultrametric $d(x,y)$ form
a finite increasing sequence: $d_{1}<d_{2}<d_{3},\:\ldots.$ Denote
by $S_{i}$, $i=1,2,\ldots$, a subset of elements $x\in\mathcal{U}$
such that $\forall x\in S_{i}$ $d(x,x_{0})=d_{i}$. We will call
the sets $S_{i},\:i=1,2,\ldots,$ ``spheres'' of radius $d_{i}$
with respect to the element $x_{0}$. For a subset consisting of a
single element $x_{0}$, we will use the notation $S_{0}$ and call
it a ``sphere'' of zero radius. Next, we will denote by $B_{i}$ a
ball of radius $d_{i}$ that contains the element $x_{0}$, i.e.,
$\forall x\in B_{i},\;d(x,x_{0})\leq d_{i}$. It is obvious that

\[
B_{i}=\bigcup_{j=0}^{i}S_{j},\;\mathcal{U}=\bigcup_{i=0}^{\infty}S_{i}.
\]
Note that, according to our conventions, $B_{0}\equiv S_{0}$.

Denote the number of points on the sphere $S_{i}$ by $M_{i}$, and
the number of points in the ball $B_{i}$ by $N_{i}$. Due to
ultrametricity, for any $x_{i}\in S_{i}$ and $x_{j}\in S_{j}$,
\[
d(x_{i},x_{j})=\max\left\{ d(x_{0},x_{i}),d(x_{0},x_{j})\right\}
=\max\left\{ d_{i},d_{j}\right\};
\]
therefore, the distance $d(x_{i},x_{j})$ does not depend on the
choice of the points $x_{i}\in S_{i}$ and $x_{j}\in S_{j}$, and we
can speak of the distance $d_{ij}$ between the spheres $S_{i}$ and
$S_{j}$, where

\begin{equation}
d_{ij}\equiv d(x_{i},x_{j})=\max\left\{ d_{i},d_{j}\right\} ,\;\forall x_{i}\in S_{i},\:\forall x_{j}\in S_{j}.\label{dij}
\end{equation}

Denote by $I_{i}\left(x\right)$ and $J_{i}\left(x\right)$ the
characteristic functions of the sphere $S_{i}$ and the ball
$B_{i}$, respectively. For any two functions $f_{1}\left(x\right)$
and $f_{2}\left(x\right)$ belonging to the linear hull of the
functions $\left\{ I_{i}\left(x\right)\right\} $,
$i=0,1,2,\ldots$, define the inner product
$\left(f_{1},f_{2}\right)=\sum_{x\in\mathcal{U}}f_{1}\left(x\right)f_{2}\left(x\right)$
and the norm $\left\Vert f_{1}\right\Vert
=\sqrt{\left(f_{1},f_{1}\right)}$. We will call all the functions
on $\mathcal{U}$ from the linear hull of the functions $\left\{
I_{i}\left(x\right)\right\} $, $i=0,1,2,\ldots$, functions
spherically symmetric with respect to the point $x_{0}$. Denote by
$F_{x_{0}}$ the class of spherically symmetric functions on
$\mathcal{U}$ that have finite norm. It is obvious that the
functions $\left\{ I_{i}\left(x\right)\right\} $ form an
orthogonal basis on $F_{x_{0}}$. Since
$I_{i}\left(x\right)=J_{i}\left(x\right)-J_{i-1}\left(x\right)$
and $I_{0}\left(x\right)\equiv
J_{0}\left(x\right)=I_{x_{0}}\left(x\right)$, the functions
$\left\{ J_{i}\left(x\right)\right\} $ also form a (nonorthogonal)
basis.

It is easily seen that $\left(\hat{A}f\right)\left(x\right)\in
F_{x_{0}}$ for any function $f\left(x\right)\in F_{x_{0}}$. To
prove this fact it suffices to verify that, for any $j$, the
action of the first term of the operator $\hat{A}$ on the function
$I_{j}\left(x\right)$ can be represented as a linear combination
of functions $\left\{ I_{i}\left(x\right)\right\} $. This is
confirmed by direct calculation with the use of (\ref{dij}):

\[
\sum_{y\in U}K\left(d\left(x,y\right)\right)\left(I_{j}\left(y\right)-I_{j}\left(x\right)\right)=
\]
\[
=\delta_{0j}\sum_{i=0}^{\infty}K\left(d_{i0}\right)I_{j}\left(x\right)-K\left(d_{j0}\right)I_{j}\left(x\right)+\sum_{k=0}^{\infty}N_{j}K\left(d_{kj}\right)I_{k}\left(x\right)-\sum_{i=0}^{\infty}N_{i}K\left(d_{ji}\right)I_{j}\left(x\right).
\]
Thus, the space $F_{x_{0}}$ is invariant with respect to the
operator on the right-hand side of equation (\ref{Re_Di_Eq}) and
is the generator of the semigroup of time translations
$f_{0}\left(x\right)\rightarrow f\left(x,t\right)$. Therefore, the
solution of the Cauchy problem for equation (\ref{Re_Di_Eq}) with
the initial condition (\ref{Re_Di_Nc}) also lies in the space
$F_{x_{0}}$ for any $t$ .

To find a solution to the Cauchy problem
(\ref{Re_Di_Eq})--(\ref{Re_Di_Nc}) we need Theorem \ref{th_1}.

\begin{theorem} Suppose that the following conditions are satisfied\textup:

\begin{equation}
\sum_{i=0}^{\infty}\dfrac{1}{N_{i}}<\infty,\;\sum_{j=1}^{\infty}e^{-\alpha d_{j}}<\infty.\label{restr}
\end{equation}
Then the functions\emph{
\begin{equation}
\phi_{i}\left(x\right)=\left(N_{i-1}\left(1-\dfrac{N_{i-1}}{N_{i}}\right)\right)^{-\tfrac{1}{2}}\left(J_{i-1}\left(x\right)-\dfrac{N_{i-1}}{N_{i}}J_{i}\left(x\right)\right),\;i=1,2,\ldots,\label{basis_f}
\end{equation}
}form an orthonormal basis in $F_{x_{0}}$ and are eigenfunctions
of the operator (\ref{Op_A}) with eigenvalues $-\lambda_{i}$,
where\emph{ }
\[
\lambda_{i}=\sum_{j=i}^{\infty}\left(K\left(d_{j}\right)-K\left(d_{j+1}\right)\right)N_{j}=
\]
\begin{equation}
=\sum_{j=i}^{\infty}e^{-\alpha d_{j}}\left(1-e^{-\alpha\left(d_{j+1}-d_{j}\right)}\dfrac{N_{j}}{N_{j+1}}\right).\label{lambda_gn}
\end{equation}

\label{th_1}

\end{theorem}

The proof of Theorem \ref{th_1} is given in Appendix 1.

Expanding $f\left(x,t\right)$ in the basis (\ref{basis_f}),
\[
f\left(x,t\right)=\sum_{i=1}^{\infty}f_{i}\left(t\right)\phi_{i}\left(x\right),
\]
 from (\ref{Re_Di_Eq}) we obtain
\[
\dfrac{df_{i}\left(t\right)}{dt}=-\lambda_{i}f_{i}\left(t\right).
\]
If the initial condition (\ref{Re_Di_Nc_gen}) lies in the class of
functions from $F_{x_{0}}$, then, taking into account its
expansion in the basis (\ref{basis_f}),
\[
f\left(x,0\right)=\sum_{j=1}^{\infty}c_{j}\phi_{j}\left(x\right),
\]
we obtain a solution of the Cauchy problem for the ultrametric
diffusion equation in the form

\begin{equation}
f\left(x,t\right)=\sum_{i=1}^{\infty}c_{i}\exp\left(-\lambda_{i}t\right)\phi_{i}\left(x\right).\label{f_x_t_gen}
\end{equation}
In particular, for the initial condition concentrated at the
center of spherical symmetry,
\begin{equation}
f\left(x,0\right)=J_{0}\left(x\right)=\sum_{j=1}^{\infty}N_{j-1}^{-\tfrac{1}{2}}\left(1-\dfrac{N_{j-1}}{N_{j}}\right)^{\tfrac{1}{2}}\phi_{j}\left(x\right),\label{Re_Di_Nc}
\end{equation}
solution (\ref{f_x_t_gen}) is rewritten as

\begin{equation}
f\left(x,t\right)=\sum_{i=1}^{\infty}N_{i-1}^{-1}\left(J_{i-1}\left(x\right)-\dfrac{N_{i-1}}{N_{i}}J_{i}\left(x\right)\right)\exp\left(-\lambda_{i}t\right).\label{f_x_t}
\end{equation}

\section{Asymptotic behavior of the solution to the problem of spherically symmetric ultrametric random walk}

\label{sec_3}

In this section we investigate the conditions under which the
asymptotic behavior of the function (\ref{f_x_t}) as
$t\rightarrow\infty$ can be represented as the product of a power
function multiplied by a bounded function that is not
infinitesimal as $t\rightarrow\infty$ and is log-periodic under
additional conditions.

Using traditional notations, below we will write
$f_{1}\left(t\right)\sim f_{2}\left(t\right)$ as
$t\rightarrow\infty$ if
$\lim_{t\rightarrow\infty}\dfrac{f_{1}\left(t\right)}{f_{2}\left(t\right)}=1$
and $f_{1}\left(t\right)=o\left(f_{2}\left(t\right)\right)$ if
$\lim_{t\rightarrow\infty}\dfrac{f_{1}\left(t\right)}{f_{2}\left(t\right)}=0$.

We consider two scenarios of asymptotic behavior of the sequence
$N_{i}$ of the number of points in ultrametric balls with a common
center $x_{0}$ and the values of the ultrametric distance $d_{i}$
from the point $x_{0}$ to the points of the sphere $S_{i}$ as
$i\rightarrow\infty$. In the first scenario, the sequences $N_{i}$
and $d_{i}$ satisfy the following conditions:

\begin{equation}
\exists\:\theta,A,A^{\prime},\:\theta>0,\:A>0,\:A^{\prime}>0:\:A<N_{i}^{-1}e^{\theta i}<A^{\prime}\:\forall i>0,\label{Cond_N_restr}
\end{equation}
\begin{equation}
\exists\:\xi,B,\:\xi>0\:B>0,:\:\left|d_{i}-\xi i\right|<B\:\forall i>0.\label{Cond_d_restr}
\end{equation}
The second scenario is a particular case of the first scenario,
and the sequences $N_{i}$ and $d_{i}$ in this scenario satisfy
more stringent conditions:

\begin{equation}
\exists\:\theta,C:\:\theta>0,\:0<C<\infty,\:\lim_{i\rightarrow\infty}N_{i}^{-1}e^{\theta i}=C,\label{Cond_N}
\end{equation}
\begin{equation}
\exists\:\xi,D:\:\xi>0,\:0\leq D<\infty,\:\lim_{i\rightarrow\infty}\left(d_{i}-\xi i\right)=D.\label{Cond_d}
\end{equation}

We will analyze the asymptotic behavior (\ref{f_x_t}) in these two
scenarios. When conditions (\ref{Cond_N})--(\ref{Cond_d}) are
satisfied, the following representation is valid:
\begin{equation}
d_{i}=\xi i+\delta_{i},\;N_{i}=C^{-1}e^{\theta i}\left(1+\varepsilon_{i}\right),\label{D_i_N_i}
\end{equation}
where
\begin{equation}
\lim_{i\rightarrow\infty}\delta_{i}=D,\:\lim_{i\rightarrow\infty}\varepsilon_{i}=0,\:\varepsilon_{i}>-1.\label{delta_epsilon}
\end{equation}
If conditions (\ref{Cond_N})--(\ref{Cond_d}) are not satisfied but
conditions (\ref{Cond_N_restr})--(\ref{Cond_d_restr}) are
satisfied, then representation (\ref{D_i_N_i}) is also valid. In
this case, the sequences $\delta_{i}$ and $\varepsilon_{i}>-1$ are
bounded, conditions (\ref{delta_epsilon}) may not hold, and the
choice of the number $C$ depends on the definition of the sequence
$\varepsilon_{i}$. Then Theorem \ref{th_1} and representation
(\ref{D_i_N_i}) imply that the eigenvalues of the operator
$\hat{A}$ can be represented as

\begin{equation}
-\lambda_{i}=-e^{-\alpha\xi i}\sum_{j=0}^{\infty}e^{-\alpha\xi j}\left[e^{-\alpha\delta_{j+i}}\left(1-e^{-\theta}e^{-\alpha\xi}e^{-\alpha\left(\delta_{j+i+1}-\delta_{j+i}\right)}\dfrac{1+\varepsilon_{j+i}}{1+\varepsilon_{j+i+1}}\right)\right].\label{expr_lambda}
\end{equation}

In what follows we need the following theorem.

\begin{theorem}

If $a_{n}$ and $b_{n}$, $n=0,1,2,\ldots$, are two infinite
positive sequences in $\mathbb{R}$ and there exist numbers $a>1$
and $b>1$ such that the sequences $a_{n}a^{n}$ and $b_{n}b^{n}$
are bounded, then the series

\begin{equation}
S\left(t\right)=\mathop{\sum}\limits _{m=0}^{\infty}a_{m}e^{-b_{m}t}\label{Ser_S}
\end{equation}
for $t\rightarrow\infty$ satisfies the property
\[
S\left(t\right)\sim t^{-\tfrac{\log a}{\log b}}f\left(t\right),
\]
where $f\left(t\right)$ is a bounded function that is not
infinitesimal as $t\rightarrow\infty$. If, in addition,

\begin{equation}
a_{n}a^{n}\sim1,\;b_{n}b^{n}\sim1\label{a_n_b_n}
\end{equation}
as $n\rightarrow\infty$, then the function $f\left(t\right)$ is
log-periodic $f\left(bt\right)=f\left(t\right)$ and has the form
\begin{equation}
f\left(t\right)=\dfrac{1}{\log b}\sum_{m=-\infty}^{\infty}\exp\left(\dfrac{2\pi im}{\log b}\log t\right)\Gamma\left(\dfrac{\log a}{\log b}-\dfrac{2\pi im}{\log b}\right).\label{S_th}
\end{equation}

\label{th_2}

\end{theorem}

Note that this theorem is a modification of Theorem 1 in
\cite{BZ_2023}, where the proof of the theorem was partly
intuitive. Therefore, in Appendix 2 we present a complete and
rigorous proof of Theorem \ref{th_2}.

The following Theorem \ref{th_3} establishes the character of the
asymptotic behavior of the function (\ref{f_x_t}) for two
scenarios of asymptotic behavior of the sequences $N_{i}$ and
$d_{i}$ as $i\rightarrow\infty$ defined by conditions
(\ref{Cond_N_restr})--(\ref{Cond_d_restr}) and
(\ref{Cond_N})--(\ref{Cond_d}), respectively.

\begin{theorem}

If $N_{i}$ and $d_{i}$ satisfy conditions
(\ref{Cond_N_restr})--(\ref{Cond_d_restr}), then the following
asymptotic behavior of the solution (\ref{f_x_t}) of the Cauchy
problem (\ref{Re_Di_Eq})--(\ref{Re_Di_Nc}) is valid as
$t\rightarrow\infty:$

\[
f\left(x,t\right)\sim t^{-\tfrac{\theta}{\alpha\xi}}g\left(t\right),\;x\in S_{k},\;\forall k\geq0,
\]
where $g\left(t\right)$ is a bounded function that is not
infinitesimal as $t\rightarrow\infty$. If, in addition, $N_{i}$
and $d_{i}$ satisfy conditions (\ref{Cond_N})--(\ref{Cond_d}),
then the function $g\left(t\right)$ is log-periodic
$g\left(e^{\alpha\xi}t\right)=g\left(t\right)$ and has the form
\[
g\left(t\right)=\cfrac{C\left(1-e^{-\theta}\right)}{\alpha\xi}\left(\dfrac{1-e^{-\theta}e^{-\alpha\xi}}{1-e^{-\alpha\xi}}e^{-\alpha D}\right)^{-\tfrac{\theta}{\alpha\xi}}
\]
\begin{equation}
\times\sum_{m=-\infty}^{\infty}\exp\left(\dfrac{2\pi im}{\alpha\xi}\log\left(\dfrac{1-e^{-\theta}e^{-\alpha\xi}}{1-e^{-\alpha\xi}}e^{-\alpha D}t\right)\right)\Gamma\left(\dfrac{\theta}{\alpha\xi}-\dfrac{2\pi im}{\alpha\xi}\right).\label{formula_th_3}
\end{equation}

\label{th_3}

\end{theorem}

Let us prove Theorem \ref{th_3}. Suppose that the scenario
(\ref{Cond_N_restr})--(\ref{Cond_d_restr}) is realized. It follows
from formulas (\ref{Cond_N_restr}) and (\ref{expr_lambda}) that
the sequences
$N_{i}^{-1}\left(1-\dfrac{N_{i}}{N_{i+1}}\right)e^{\theta i}$ and
$\lambda_{i}e^{\alpha\xi i}$ are bounded. Then the application of
the first assertion of Theorem \ref{th_2} to the solution of
ultrametric diffusion equation (\ref{f_x_t}) expressed as

\begin{equation}
f\left(x,t\right)=f\left(x_{0},t\right)-\sum_{i=1}^{k}N_{i-1}^{-1}\left(1-\dfrac{N_{i-1}}{N_{i}}\right)\exp\left(-\lambda_{i}t\right)-\dfrac{1}{N_{k}}\exp\left(-\lambda_{k}t\right),\;x\in S_{k},\:k>0,\label{f_x_t_form}
\end{equation}
where

\[
f\left(x_{0},t\right)=\sum_{i=1}^{\infty}N_{i-1}^{-1}\left(1-\dfrac{N_{i-1}}{N_{i}}\right)\exp\left(-\lambda_{i}t\right),
\]
gives the first assertion of Theorem \ref{th_3}.

Now, suppose that scenario (\ref{Cond_N})--(\ref{Cond_d}) is
realized. In this case, the sequences $\delta_{i}$ and
$\varepsilon_{i}$ in the representations (\ref{D_i_N_i}) and
(\ref{expr_lambda}) satisfy conditions (\ref{delta_epsilon}).
Consider the limit
$\lim_{i\rightarrow\infty}\lambda_{i}e^{\alpha\xi i}$. Since the
expression in square brackets in (\ref{expr_lambda}) is bounded,
the sum in (\ref{expr_lambda}) converges uniformly in $i$, and the
limit can be taken under the summation sign. As a result, we
obtain

\begin{equation}
\lim_{i\rightarrow\infty}\lambda_{i}e^{\alpha\xi i}=e^{-\alpha D}\sum_{j=0}^{\infty}e^{-\alpha\xi j}\left(1-e^{-\theta}e^{-\alpha\xi}\right)=\eta e^{-\alpha D},\;\eta=\dfrac{1-e^{-\theta}e^{-\alpha\xi}}{1-e^{-\alpha\xi}}.\label{condition_lam}
\end{equation}
Next, from (\ref{Cond_N}) we have
\begin{equation}
\lim_{i\rightarrow\infty}N_{i}^{-1}\left(1-\dfrac{N_{i}}{N_{i+1}}\right)e^{\theta i}=C\left(1-e^{-\theta}\right).\label{condition_N}
\end{equation}
With regard to formulas (\ref{condition_lam}) and
(\ref{condition_N}), the application of the second assertion of
Theorem \ref{th_2} to function (\ref{f_x_t_form}) yields the
second assertion of Theorem \ref{th_3}.

\section{Problem of spherically symmetric ultrametric random walk
with a reaction sink }

\label{sec_4}

In this section we consider the Cauchy problem on $\mathcal{U}$
for the ultrametric diffusion equation with a reaction sink
concentrated at the point $x_{0}$,
\begin{equation}
\dfrac{df\left(x,t\right)}{dt}=\left(\hat{A}f\right)\left(x,t\right)-kJ_{0}\left(x\right)f\left(x,t\right),\label{Re_Di_Eq_S}
\end{equation}
with the initial condition of the general form
\begin{equation}
f\left(x,0\right)=f_{0}\left(x\right)\in F_{x_{0}}.\label{Ic_S}
\end{equation}
The operator on the right-hand side of equation (\ref{Re_Di_Eq_S})
does not take the functions from the class $F_{x_{0}}$. We will
seek a solution $f\left(x,t\right)$ to the Cauchy problem
(\ref{Re_Di_Eq_S})--(\ref{Ic_S}) in the class of functions
$f\left(x,t\right)\in F_{x_{0}}$ for any $t\in\mathbb{R}_{+}$ and
$\left|f\left(x,t\right)\right|<M\exp\left(at\right)$ for some
$M,\:a\in\mathbb{R}_{+},$ and any $x\in\mathcal{U}$. Passing to
the Laplace transform $\widetilde{f}\left(x,s\right)$ of the
function $f\left(x,t\right)$, we rewrite equation
(\ref{Re_Di_Eq_S}) as

\begin{equation}
s\widetilde{f}\left(x,s\right)=f_{0}\left(x\right)+\left(\hat{A}\widetilde{f}\right)\left(x,s\right)-kJ_{0}\left(x\right)\widetilde{f}\left(x,s\right).\label{Eq_Lap}
\end{equation}
The expansion of the functions $J_{0}\left(x\right)$,
$\widetilde{f}\left(x,s\right)$, and $f_{0}\left(x\right)$ in the
basis (\ref{basis_f})
\[
J_{0}\left(x\right)=\sum_{k=1}^{\infty}N_{k-1}^{-\tfrac{1}{2}}\left(1-\dfrac{N_{k-1}}{N_{k}}\right)^{\tfrac{1}{2}}\phi_{k}\left(x\right),
\]
\begin{equation}
\widetilde{f}\left(x,s\right)=\sum_{j=1}^{\infty}\widetilde{f}_{j}\left(s\right)\phi_{j}\left(x\right),\;f_{0}\left(x\right)=\sum_{j=1}^{\infty}c_{j}\phi_{j}\left(x\right)\label{decomp_basis}
\end{equation}
followed by the substitution into (\ref{Eq_Lap}) leads to the
following equation:

\[
s\sum_{j=1}^{\infty}\widetilde{f}_{j}\left(s\right)\phi_{j}\left(x\right)=\sum_{j=1}^{\infty}c_{j}\phi_{j}\left(x\right)+\sum_{j=1}^{\infty}\lambda_{j}\widetilde{f}_{j}\left(s\right)\phi_{j}\left(x\right)
\]
\begin{equation}
-k\sum_{k=1}^{\infty}N_{k-1}^{-\tfrac{1}{2}}\left(1-\dfrac{N_{k-1}}{N_{k}}\right)^{\tfrac{1}{2}}\phi_{k}\left(x\right)\sum_{j=0}^{\infty}\widetilde{f}_{j}\left(s\right)\phi_{j}\left(x\right).\label{eq_lap_bas}
\end{equation}
The following formula, which is easily verified by direct
calculation, will be useful for further transformations:

\[
\sum_{x\in U}\phi_{i}\left(x\right)\phi_{j}\left(x\right)\phi_{k}\left(x\right)=
\]
\[
=\delta_{ij}\delta_{jk}N_{k-1}^{-\tfrac{1}{2}}\left(1-\dfrac{N_{k-1}}{N_{k}}\right)^{-\tfrac{3}{2}}\left(1-3\dfrac{N_{k-1}}{N_{k}}+2\dfrac{N_{k-1}^{2}}{N_{k}^{2}}\right)+
\]
\[
+\delta_{ij}\delta_{i<k}N_{k-1}^{-\tfrac{1}{2}}\left(1-\dfrac{N_{k-1}}{N_{k}}\right)^{\tfrac{1}{2}}+
\]
\[
+\delta_{ik}\delta_{i<j}N_{j-1}^{-\tfrac{1}{2}}\left(1-\dfrac{N_{j-1}}{N_{j}}\right)^{\tfrac{1}{2}}+
\]
\begin{equation}
+\delta_{jk}\delta_{j<i}N_{i-1}^{-\tfrac{1}{2}}\left(1-\dfrac{N_{i-1}}{N_{i}}\right)^{\tfrac{1}{2}}\;\forall i,j,k,\label{phi_phi_phi}
\end{equation}
where we used the notation
\[
\delta_{j<i}=\begin{cases}
1, & j<i,\\
0, & j\geq i.
\end{cases}
\]
Multiplying (\ref{eq_lap_bas}) by $\phi_{i}\left(x\right)$,
summing over $x\in\mathcal{U}$, and using (\ref{phi_phi_phi}), we
obtain the following equations for the functions
$\widetilde{f}_{i}\left(s\right)$:

\[
s\widetilde{f}_{i}\left(s\right)=c_{i}-\lambda_{i}\widetilde{f}_{i}\left(s\right)-
\]

\[
-kN_{i-1}^{-\tfrac{1}{2}}\left(1-\dfrac{N_{i-1}}{N_{i}}\right)^{\tfrac{1}{2}}\widetilde{f}_{0}\left(s\right)N^{-\tfrac{1}{2}}-kN_{i-1}^{-\tfrac{1}{2}}\left(1-\dfrac{N_{i-1}}{N_{i}}\right)^{\tfrac{1}{2}}\sum_{j=1}^{\infty}N_{j-1}^{-\tfrac{1}{2}}\left(1-\dfrac{N_{j-1}}{N_{j}}\right)^{\tfrac{1}{2}}\widetilde{f}_{j}\left(s\right).
\]
Taking into account that

\begin{equation}
\widetilde{f}\left(x_{0},s\right)=\sum_{j=1}^{\infty}N_{j-1}^{-\tfrac{1}{2}}\left(1-\dfrac{N_{j-1}}{N_{j}}\right)^{\tfrac{1}{2}}\widetilde{f}_{j}\left(s\right),\label{til_f_x_0_s}
\end{equation}
we can rewrite the last equation as

\begin{equation}
\widetilde{f}_{i}\left(s\right)=\dfrac{c_{i}}{s+\lambda_{i}}-N_{i-1}^{-\tfrac{1}{2}}\left(1-\dfrac{N_{i-1}}{N_{i}}\right)^{\tfrac{1}{2}}\dfrac{k\widetilde{f}\left(x_{0},s\right)}{s-\lambda_{i}},\;i>0.\label{f_i_s}
\end{equation}
The multiplication of (\ref{f_i_s}) by
$N_{i-1}^{-\tfrac{1}{2}}\left(1-\dfrac{N_{i-1}}{N_{i}}\right)^{\tfrac{1}{2}}$
followed by the summation over $i$ yields an equation for the
function $\widetilde{f}\left(x_{0},s\right)$, which, upon
introducing the notations

\[
J\left(s\right)=\sum_{j=1}^{\infty}N_{j-1}^{-1}\left(1-\dfrac{N_{j-1}}{N_{j}}\right)\dfrac{1}{s+\lambda_{j}},
\]

\[
C\left(s\right)=\sum_{j=1}^{\infty}N_{j-1}^{-\tfrac{1}{2}}\left(1-\dfrac{N_{j-1}}{N_{j}}\right)^{\tfrac{1}{2}}\dfrac{c_{i}}{s+\lambda_{j}},
\]
can be expressed as
\[
\widetilde{f}\left(x_{0},s\right)=C\left(s\right)-k\widetilde{f}\left(x_{0},s\right)J\left(s\right).
\]
The last equation can be rewritten as
\begin{equation}
\widetilde{f}\left(x_{0},s\right)=\dfrac{C\left(s\right)}{1+kJ\left(s\right)}.\label{f_0(s)}
\end{equation}
Taking into account (\ref{f_i_s}) and (\ref{f_0(s)}), we write a
solution for the Laplace transforms
$\widetilde{f}_{i}\left(s\right)$ of the expansion coefficients of
$f\left(x,t\right)$ in the basis (\ref{basis_f}) as

\begin{equation}
\widetilde{f}_{i}\left(s\right)=\dfrac{c_{i}}{s+\lambda_{i}}-N_{i-1}^{-\tfrac{1}{2}}\left(1-\dfrac{N_{i-1}}{N_{i}}\right)^{\tfrac{1}{2}}\dfrac{1}{s-\lambda_{i}}\dfrac{kC\left(s\right)}{1+kJ\left(s\right)},\;i>0.\label{f_i}
\end{equation}
In this case, the Laplace transform
$\widetilde{f}\left(x,s\right)$ of the solution to the Cauchy
problem for equation (\ref{Re_Di_Eq_S})--(\ref{Ic_S}) has the
form:
\[
\widetilde{f}\left(x,s\right)=\sum_{j=1}^{\infty}\widetilde{f}_{j}\left(s\right)\phi_{j}\left(x\right)=\sum_{j=1}^{\infty}\dfrac{c_{j}}{s+\lambda_{j}}\phi_{j}\left(x\right)
\]

\begin{equation}
-\sum_{j=1}^{\infty}N_{j-1}^{-\tfrac{1}{2}}\left(1-\dfrac{N_{j-1}}{N_{j}}\right)^{\tfrac{1}{2}}\dfrac{1}{s+\lambda_{j}}\dfrac{kC\left(s\right)}{1+kJ\left(s\right)}\phi_{j}\left(x\right).\label{gen_solution}
\end{equation}
Let us also find the Laplace transform
$\tilde{S}\left(s\right)=\sum_{x\in\mathcal{U}}\widetilde{f}\left(x,s\right)$
of the probability measure
$S\left(t\right)=\sum_{x\in\mathcal{U}}f\left(x,t\right)$ of the
distribution $f\left(x,t\right)$ on the whole space $\mathcal{U}$.
The summation of equation (\ref{Re_Di_Eq_S}) over
$x\in\mathcal{U}$ yields
\[
\dfrac{dS\left(t\right)}{dt}=-kf\left(x_{0},t\right)=0.
\]
Passing to the Laplace transforms $\widetilde{S}\left(s\right)$
and $\widetilde{f}\left(x_{0},s\right)$ of the functions
$S\left(t\right)$ and $f\left(x_{0},t\right)$ and taking into
account that $S\left(0\right)=1$, we obtain
\[
s\widetilde{S}\left(s\right)=1-k\widetilde{f}\left(x_{0},s\right);
\]
hence

\begin{equation}
\tilde{S}\left(s\right)=\dfrac{1}{s}\left(1-k\widetilde{f}\left(x_{0},s\right)\right).\label{S_f_0}
\end{equation}

\section{Asymptotic behavior of the solution to the problem of spherically symmetric
ultrametric random walk with a reaction sink}

\label{sec_5}

In this section we establish conditions under which the asymptotic
behavior of the function $S\left(t\right)$ as $t\rightarrow\infty$
has the form of a power function multiplied by a bounded function,
as well as conditions under which this function is log-periodic.
Take the initial condition (\ref{Ic_S}) of the Cauchy problem for
equation (\ref{Re_Di_Eq_S}) in the form (\ref{Re_Di_Nc}). In this
case, we have

\[
C\left(s\right)=J\left(s\right)=\sum_{i=1}^{\infty}N_{i-1}^{-1}\left(1-\dfrac{N_{i-1}}{N_{i}}\right)\dfrac{1}{s+\lambda_{i}};
\]
then the Laplace transform (\ref{f_0(s)}) of the solution at the
point $x_{0}$ is expressed as

\begin{equation}
\widetilde{f}\left(x_{0},s\right)=\dfrac{J\left(s\right)}{1+kJ\left(s\right)}=\dfrac{1}{k}\left(1-\dfrac{1}{1+kJ\left(s\right)}\right).\label{f_0(s)_sink}
\end{equation}
The function $J\left(s\right)$ has simple poles at the points
$s=-\lambda_{i}$. Since
\[
\lim_{s\rightarrow\lambda_{i}}\widetilde{f}\left(x_{0},s\right)=k^{-1},
\]
it follows that the points $s=-\lambda_{i}$ are points of
removable discontinuity for the function
$\widetilde{f}\left(x_{0},s\right)$, and the function
$\widetilde{f}\left(x_{0},s\right)$ can be redefined at these
points. After such a redefinition, the function
(\ref{f_0(s)_sink}) will be holomorphic in the entire expanded
complex plane except for the points at which this function has
simple poles defined by the equation
\begin{equation}
1+kJ\left(s\right)=0.\label{eq_pol}
\end{equation}
It can be shown by a graphical analysis of equation (\ref{eq_pol})
that its roots are $s=-\nu_{i}\;i=1,2,...$, where the numbers
$\nu_{i}$ satisfy the inequalities $\lambda_{1}<\nu_{1}$ and
$\lambda_{j`}<\nu_{j}<\lambda_{j-1}$ for $j>1$. For further
analysis it is convenient to use the following parameterization of
the numbers $\nu_{i}$:

\begin{equation}
\nu_{i}=\lambda_{i}\left(1+\Delta_{i}\right),\label{nu_lambda_delta}
\end{equation}
where
\begin{equation}
0<\Delta_{1},\;0<\Delta_{j}<\dfrac{\lambda_{j-1}}{\lambda_{j`}}-1,\;j>1.\label{Delta_restr}
\end{equation}
The function $\widetilde{f}\left(x_{0},s\right)$ is not
meromorphic since $s=0$ is an essentially singular point at which
the poles are concentrated. Nevertheless, the function
$H\left(z\right)=\widetilde{f}\left(x_{0},\dfrac{1}{z}\right)$,
where $z=\dfrac{1}{s}$, is meromorphic for $z\in\mathbb{C}$, and
${\displaystyle
\lim_{z\rightarrow\infty}H\left(z\right)=\dfrac{1}{k}}$,
${\displaystyle \lim_{z\rightarrow0}H\left(z\right)=0}$ in view of
the fact that ${\displaystyle
\lim_{s\rightarrow0,\:\mathrm{Re}s>0}\widetilde{f}\left(x_{0},s\right)=\dfrac{1}{k}}$
and ${\displaystyle
\lim_{s\rightarrow\infty}\widetilde{f}\left(x_{0},s\right)=0}$.
Since $|H\left(z\right)|\leq A|z|^{m}$ as $z\rightarrow\infty$, it
follows from Mittag--Leffler's expansion theorem that
$H\left(z\right)$ can be represented as a uniformly converging
(except for a countable number of simple poles) series
$H\left(z\right)={\displaystyle
\sum_{i=1}^{\infty}\dfrac{a_{k}}{z+\frac{1}{\nu_{i}}}}$.
Therefore, the function
$\widetilde{f}\left(x_{0},s\right)=H\left(\dfrac{1}{s}\right)$ can
also be represented as a uniformly converging series
\begin{equation}
\widetilde{f}\left(x_{0},s\right)={\displaystyle \sum_{i=1}^{\infty}\dfrac{b_{i}}{s+\nu_{i}},}\label{F_poles}
\end{equation}
where $b_{i}$ are the residues of the function
$\widetilde{f}\left(x_{0},s\right)$ at the poles $s=-\nu_{i}$:

\begin{equation}
b_{i}={\displaystyle \underset{s=-\nu_{i}}{\mathrm{Res}\widetilde{f}\left(x_{0},s\right)}}=\dfrac{1}{k^{2}\left|J^{\prime}\left(-\nu_{i}\right)\right|}.\label{b_k}
\end{equation}
Since the function $f\left(x_{0},t\right)$ is the inverse Laplace
transform of the meromorphic function
$\widetilde{f}\left(x_{0},s\right)$, it can be represented as an
infinite series:
\begin{equation}
f\left(x_{0},t\right)={\displaystyle \sum_{i=1}^{\infty}}b_{i}\exp\left(-\nu_{i}t\right)\risingdotseq\widetilde{f}\left(x_{0},s\right)={\displaystyle \sum_{i=1}^{\infty}\dfrac{b_{i}}{s+\nu_{i}}.}\label{f_x_0_t}
\end{equation}
Taking into account (\ref{S_f_0}), we have
\[
\tilde{S}\left(s\right)=\dfrac{1}{s}\left(1-k\sum_{i=1}^{\infty}\dfrac{b_{i}}{s+\nu_{i}}\right)=\dfrac{1}{s}\left(1-k\sum_{i=1}^{\infty}\dfrac{b_{i}}{\nu_{i}}\right)+k\sum_{i=1}^{\infty}\dfrac{1}{s+\nu_{i}}\dfrac{b_{i}}{\nu_{i}}.
\]
It follows from the property ${\displaystyle
\widetilde{f}\left(x_{0},0\right)=\dfrac{1}{k}}$ that
$\sum_{i=1}^{\infty}\dfrac{b_{i}}{\nu_{i}}=\dfrac{1}{k}$; hence
$\tilde{S}\left(s\right)$ takes the form
\[
\tilde{S}\left(s\right)=k\sum_{i=1}^{\infty}\dfrac{1}{s+\nu_{i}}\dfrac{b_{i}}{\nu_{i}}.
\]
The last formula implies the following representation for
$S\left(t\right)$:

\begin{equation}
S\left(t\right)=k\sum_{i=1}^{\infty}\dfrac{b_{i}}{\nu_{i}}\exp\left(-\nu_{i}t\right).\label{S(t)_sum}
\end{equation}

Next, we formulate Theorem \ref{th_4}, which establishes a
sufficient condition for the asymptotics of the function
(\ref{S(t)_sum}) as $t\rightarrow\infty$ to have the form of a
power function multiplied by a bounded function, which is
log-periodic under certain conditions.

\begin{theorem}

If $\alpha>\dfrac{\theta}{\xi}$ and conditions
(\ref{Cond_N_restr}) and (\ref{Cond_d_restr}) are satisfied, then
the function $S\left(t\right)$ has the following asymptotic
behavior as $t\rightarrow\infty:$
\[
S\left(t\right)\sim t^{-\tfrac{\alpha\xi-\theta}{\alpha\xi}}h\left(t\right),
\]
where $h\left(t\right)$ is a bounded function that is not
infinitesimal as $t\rightarrow\infty$. If, in addition, conditions
(\ref{Cond_N}) and (\ref{Cond_d}) are satisfied, then there exists
$\lim_{i\rightarrow\infty}\Delta_{i}=\Delta>0$, and the function
$h\left(t\right)$ is log-periodic
$h\left(e^{\alpha\xi}t\right)=h\left(t\right)$ of the form
\[
h\left(t\right)=\dfrac{\Delta^{2}e^{-\theta}\eta e^{-\alpha D}}{k^{2}C\left(1+\Delta\right)\left(1-e^{-\theta}\right)}\dfrac{1}{\alpha\xi}\left(\dfrac{1-e^{-\theta}e^{-\alpha\xi}}{1-e^{-\alpha\xi}}e^{-\alpha D}\left(1+\Delta\right)\right)^{-\tfrac{\alpha\xi-\theta}{\alpha\xi}}
\]

\[
\times\sum_{m=-\infty}^{\infty}\exp\left(\dfrac{2\pi im}{\alpha\xi}\log\left(\dfrac{1-e^{-\theta}e^{-\alpha\xi}}{1-e^{-\alpha\xi}}e^{-\alpha D}\left(1+\Delta\right)t\right)\right)\Gamma\left(\dfrac{\alpha\xi-\theta}{\alpha\xi}-\dfrac{2\pi im}{\alpha\xi}\right).
\]
.

\label{th_4}

\end{theorem}

The proof of Theorem \ref{th_4} is given in Appendix 3.

\section{Discussion}

In the present study we have obtained analytic solutions to the
Cauchy problem for an ultrametric diffusion equation on an
arbitrary countable ultrametric space with an initial condition
spherically symmetric with respect to some point $x_{0}$. These
solutions have been obtained both for the equation of pure
ultrametric diffusion and for the ultrametric diffusion equation
with a reaction sink. Using these results, we have investigated
the large-time asymptotics of the solutions and found conditions
under which these asymptotics have the form of a power law, with
negative exponent, modulated by a bounded function that does not
tend to zero. These conditions are imposed on the sequence $N_{i}$
of the number of points on ultrametric spheres with center at the
point $x_{0}$, as well as on the sequence $d_{i}$ of ultrametric
distances from the point $x_{0}$ to the points of the $i$th
sphere, and have the form
(\ref{Cond_N_restr})--(\ref{Cond_d_restr}). We have also shown
that, under additional constraints (\ref{Cond_N})--(\ref{Cond_d})
on $N_{i}$ and $d_{i}$, the function modulating the power
asymptotics is log-periodic. These results are formulated in the
form of Theorems \ref{th_3} and \ref{th_4}. Let us discuss some
important consequences of these results.

In spite of the fact that the formulations of Theorems \ref{th_3}
and \ref{th_4} contain the solution of the Cauchy problem with a
special initial condition concentrated at the center of spherical
symmetry, it is easy to understand that these theorems are also
valid for any (not necessarily spherically symmetric) initial
condition whose support is any compact set $A$ belonging to some
ultrametric ball of fixed radius $d_{k}$. The explanation is given
by the fact that in this case the time dependence of the
distribution function in all ultrametric spheres of radius $d_{i}$
with respect to any point $x_{0}\in A$ for $i>k$ is the same as
the time dependence of the distribution function in the Cauchy
problem with the initial condition concentrated at the point
$x_{0}$, and hence the asymptotic behavior of the solution is not
changed. For this reason, both conditions
(\ref{Cond_N_restr})--(\ref{Cond_d_restr}) and conditions
(\ref{Cond_N})--(\ref{Cond_d}) are sufficiently general and valid
for an initial condition with support on any compact subset of
ultrametric space.

Note also that the presence of logarithmic oscillations in the
asymptotics of the solutions of ultrametric diffusion equations
under conditions (\ref{Cond_N})--(\ref{Cond_d}) can be understood
on a purely intuitive basis from the requirement of discrete scale
invariance (see, for example, \cite{BZ_2023}). The solution of
this equation in the class of functions spherically symmetric with
respect to some point $x_{0}$ can be formally represented as a
series of weighted exponentials of the form
$\varphi\left(t\right)=\sum_{i}a_{i}\exp\left(-b_{i}t\right)$,
where $-b_{i}<0$ are the eigenvalues of the operator on the
right-hand side of the ultrametric diffusion equation and the
coefficients $a_{i}>0$ are determined by the initial conditions of
the Cauchy problem. The asymptotic behavior of the series
$\varphi\left(t\right)$ in the form of a power law modulated by
logarithmic oscillations is a consequence of the asymptotic
discrete scale invariance of the function $\varphi\left(t\right)$.
The asymptotic discrete scale invariance of
$\varphi\left(t\right)$ implies that there exist numbers $k$ and
$\beta$ such that the relation $\varphi\left(kt\right)\sim
k^{\beta}\varphi\left(t\right)\left(1+o\left(1\right)\right)$
holds as $t\rightarrow\infty$. This relation requires, in turn,
that the following conditions should be imposed on the
coefficients $a_{i}$ and $b_{i}$: $\dfrac{b_{i+1}}{b_{i}}\sim k$
and $\dfrac{a_{i+1}}{a_{i}}\sim k^{-\beta}$ as
$i\rightarrow\infty$. For the initial condition with support on a
compact set, the asymptotic behavior of the coefficients $a_{i}$
as $i\rightarrow\infty$ is completely determined by the asymptotic
behavior of the number of points $N_{i}$ in spherically symmetric
ultrametric balls $B_{i}$, which leads to an asymptotic
exponential dependence of $N_{i}$ on $i$ (condition
(\ref{Cond_N})). In turn, the asymptotic behavior of the
coefficients $b_{i}$ as $i\rightarrow\infty$ is mainly determined
by the asymptotic behavior of the sequence $d_{i}$ of values of
the ultrametric distance between the point $x_{0}$ and the points
of the sphere $S_{i}$ and leads to an asymptotic linear dependence
of $d_{i}$ on $i$ (the condition in (\ref{Cond_d})).

Conditions (\ref{Cond_N})--(\ref{Cond_d}), which guarantee the
presence of modulating logarithmic oscillations in the asymptotics
of the solutions, are rather stringent and strongly restrict the
class of ultrametric spaces and the class of ultrametrics.
According to the structure of the proof of Theorems \ref{th_3} and
\ref{th_4}, any relaxation of these conditions leads to the
violation of the logic of the proof of these theorems and, as a
consequence, to their failure. Nevertheless, we stress that,
according to the logical structure of the proof of Theorems
\ref{th_3} and \ref{th_4}, conditions
(\ref{Cond_N})--(\ref{Cond_d}) are only sufficient conditions for
the given asymptotic behavior, and we cannot assert that these
conditions are necessary.

Note that it is the description of power relaxation laws directly
observed in experiments that is important in the simulation of
disordered systems such as proteins. As pointed out in
\cite{BZ_2023}, at present there appear to be no clear experiments
with disordered systems in which one could draw reliable
conclusions about the existence of purely logarithmic
oscillations. In the present study we have shown that the main
attribute of power relaxation laws in ultrametric random walk
models is the modulation of these laws by some bounded function
that does not tend asymptotically to zero and should not
necessarily satisfy the aperiodicity condition. In view of the
possible application of these models to the description of
relaxation dynamics of protein and other disordered systems, it is
important to reveal possible modulations of power laws observed in
experiments with such systems.

\section*{Acknowledgments}

The study was supported by the Ministry of Higher
Education and Science of Russia by the State assignment to
educational and research institutions under project no.
FSSS-2023-0009.

\section*{Data Availability Statement}

The data supporting the findings of this study are available within
the article and its supplementary material. All other relevant source
data are available from the corresponding author upon reasonable request.

\section*{Appendix 1. Proof of Theorem \ref{th_1}}

\label{App_1}

In this appendix we present the proof of Theorem \ref{th_1}. The
orthonormality of the set of functions (\ref{basis_f})
\[
\left(\phi_{i},\phi_{j}\right)=\sum_{x\in U}\phi_{i}\left(x\right)\phi_{j}\left(x\right)=\delta_{ij}
\]
is proved by direct calculation. Let us prove that the set of
functions (\ref{basis_f}) forms a basis in $F$. Since the
functions $\left\{ J_{i}\left(x\right)\right\} $ form a basis in
$F$, it suffices to show that for any $i$ the function
$J_{i}\left(x\right)$ can be expanded in terms of the functions
(\ref{basis_f}). Let us show that the following squared norm
vanishes under condition (\ref{restr}):

\[
\left\Vert J_{i}\left(x\right)-N_{i}\sum_{j=i+1}^{\infty}N_{j-1}^{-\tfrac{1}{2}}\left(1-\dfrac{N_{j-1}}{N_{j}}\right)^{\tfrac{1}{2}}\phi_{j}\left(x\right)\right\Vert ^{2}=
\]
\[
=\sum_{x\in U}J_{i}\left(x\right)-2N_{i}\sum_{j=i+1}^{\infty}N_{j-1}^{-\tfrac{1}{2}}\left(1-\dfrac{N_{j-1}}{N_{j}}\right)^{\tfrac{1}{2}}\sum_{x\in U}J_{i}\left(x\right)\phi_{j}\left(x\right)+
\]
\[
+N_{i}^{2}\sum_{j=i+1}^{\infty}N_{j-1}^{-1}\left(1-\dfrac{N_{j-1}}{N_{j}}\right)\sum_{x\in U}\left(\phi_{j}\left(x\right)\right)^{2}=
\]
\[
=N_{i}-2N_{i}\sum_{j=i+1}^{\infty}N_{j-1}^{-\tfrac{1}{2}}\left(1-\dfrac{N_{j-1}}{N_{j}}\right)^{\tfrac{1}{2}}N_{j-1}^{-\tfrac{1}{2}}N_{i}\left(1-\dfrac{N_{j-1}}{N_{j}}\right)^{\tfrac{1}{2}}+
\]
\[
+N_{i}^{2}\sum_{j=i+2}^{\infty}N_{j-1}^{-1}\left(1-\dfrac{N_{j-1}}{N_{j}}\right)=
\]
\[
=N_{i}-N_{i}^{2}\left(\sum_{j=i+1}^{\infty}N_{j-1}^{-1}-\sum_{j=i+2}^{\infty}N_{j}^{-1}\right)=0
\]
Hence,
\begin{equation}
J_{i}\left(x\right)=N_{i}\sum_{j=i+1}^{\infty}N_{j-1}^{-\tfrac{1}{2}}\left(1-\dfrac{N_{j-1}}{N_{j}}\right)^{\tfrac{1}{2}}\phi_{j}\left(x\right)\label{Theorem_1}
\end{equation}
in the space $F_{x_{0}}$, which proves the first assertion of
Theorem \ref{th_1}.

Next, we calculate the action of the operator (\ref{Op_A}) on the
function
$J_{i-1}\left(x\right)=\sum_{k=0}^{i-1}I_{k}\left(x\right)$,
$i>0$. If $x\notin B_{i-1}$, then
\[
\left(\hat{A}J_{i-1}\right)\left(x\right)=\sum_{y\in U}K\left(d\left(x,y\right)\right)J_{i-1}\left(y\right)=\sum_{j=0}^{i-1}\sum_{y\in S_{j}}K\left(d\left(x,y\right)\right)J_{i-1}\left(y\right)=
\]

\[
=\sum_{j=0}^{i-1}\sum_{k=i}^{\infty}M_{j}I_{k}\left(x\right)K\left(d_{k}\right)=N_{i-1}\sum_{k=i}^{\infty}K\left(d_{k}\right)I_{k}\left(x\right).
\]
If $x\in B_{i-1}$, then

\[
\left(\hat{A}J_{i-1}\right)\left(x\right)=-\sum_{y\in U}\left(1-J_{i-1}\left(y\right)\right)K\left(d\left(x,y\right)\right)=
\]
\[
=-\sum_{j=i}^{\infty}\sum_{y\in S_{j}}K\left(d\left(x,y\right)\right)J_{i-1}\left(x\right)=-\sum_{j=i}^{\infty}M_{j}K\left(d_{j}\right)J_{i-1}\left(x\right)=
\]
\[
=-\sum_{j=i}^{\infty}\left(N_{j}-N_{i-1}\right)K\left(d_{j}\right)J_{i-1}\left(x\right).
\]
Thus,

\[
\left(\hat{A}J_{i-1}\right)\left(x\right)=N_{i-1}\sum_{j=i}^{\infty}K\left(d_{j}\right)I_{j}\left(x\right)-\sum_{j=i}^{\infty}\left(N_{j}-N_{j-1}\right)K\left(d_{j}\right)J_{i-1}\left(x\right).
\]
Next, for the function
$\dfrac{1}{N_{i-1}}J_{i-1}\left(x\right)-\dfrac{1}{N_{i}}J_{i}\left(x\right)$
we have
\[
\hat{A}\left(\dfrac{1}{N_{i-1}}J_{i-1}\left(x\right)-\dfrac{1}{N_{i}}J_{i}\left(x\right)\right)=
\]
\[
=\sum_{j=i}^{\infty}K\left(d_{j}\right)I_{j}\left(x\right)-\dfrac{1}{N_{i-1}}\sum_{j=i}^{\infty}N_{j}K\left(d_{j}\right)J_{i-1}\left(x\right)+\dfrac{1}{N_{i-1}}\sum_{j=i}^{\infty}N_{j-1}K\left(d_{j}\right)J_{i-1}\left(x\right)-
\]
\[
-\sum_{j=i+1}^{\infty}K\left(d_{j}\right)I_{j}\left(x\right)+\dfrac{1}{N_{i}}\sum_{j=i+1}^{\infty}N_{j}K\left(d_{j}\right)J_{i}\left(x\right)-\dfrac{1}{N_{i}}\sum_{j=i+1}^{\infty}N_{j-1}K\left(d_{j}\right)J_{i}\left(x\right)=
\]

\[
=-\dfrac{1}{N_{i-1}}\sum_{j=i}^{\infty}N_{j}K\left(d_{j}\right)J_{i-1}\left(x\right)+\dfrac{1}{N_{i-1}}\sum_{j=i+1}^{\infty}N_{j-1}K\left(d_{j}\right)J_{i-1}\left(x\right)-
\]
\[
+\dfrac{1}{N_{i}}\sum_{j=i}^{\infty}N_{j}K\left(d_{j}\right)J_{i}\left(x\right)-\dfrac{1}{N_{i}}\sum_{j=i+1}^{\imath}N_{j-1}K\left(d_{j}\right)J_{i}\left(x\right)=
\]
\[
=-\left(\sum_{j=i}^{\infty}N_{j}\left(K\left(d_{j}\right)-K\left(d_{j+1}\right)\right)\right)\left(\dfrac{1}{N_{i-1}}J_{i-1}\left(x\right)-\dfrac{1}{N_{i}}J_{i}\left(x\right)\right),
\]
which proves Theorem \ref{th_1}.

\section*{Appendix 2. Proof of Theorem \ref{th_2}}

\label{App_2}

In this appendix we present the proof of Theorem \ref{th_2}.
First, we prove Lemma \ref{lemma_1}.

\begin{lemma}

\label{lemma_1}

Let
\[
s\left(x\right)=\sum_{n=-\infty}^{\infty}c_{n}\left(x\right)\exp\left(2\pi
inx\right)
\]
be a series such that for all $x\in\mathbb{R}$ the series
$\sum_{n=-\infty}^{\infty}\left|c_{n}\left(x\right)\right|$
converges and the functions $c_{n}\left(x\right)$ are not
infinitesimal as $x\rightarrow\infty$ simultaneously for all $n$.
Then the function $s\left(x\right)$ is bounded and is not
infinitesimal as $x\rightarrow\infty$.

\end{lemma}

To prove Lemma \ref{lemma_1}, notice that the function
$c_{n}\left(x\right)$ can be represented on the interval
$\left[x_{0},x_{0}+1\right]$ as a sum of a linear function and a
function that takes the same values at the ends of the interval:

\[
c_{n}\left(x\right)=\alpha_{n}\left(x_{0}\right)x+\left(c_{n}\left(x\right)-\alpha_{n}\left(x_{0}\right)x\right),\;\alpha_{n}\left(x_{0}\right)=c_{n}\left(x_{0}+1\right)-c_{n}\left(x_{0}\right).
\]
Then the following Fourier series representation is valid on this
interval:

\begin{equation}
c_{n}\left(x\right)=\alpha_{n}\left(x_{0}\right)x+\sum_{m=-\infty}^{\infty}a_{nm}\left(x_{0}\right)\exp\left(2\pi im\left(x-x_{0}\right)\right),\;a_{nm}\left(x_{0}\right)=a_{n\left(-m\right)}^{\ast}\left(x_{0}\right).\label{c_n}
\end{equation}
Using expansion (\ref{c_n}), we represent $s\left(x\right)$ as
\begin{equation}
s\left(x\right)=x\sum_{k=-\infty}^{\infty}\alpha_{k}\left(x_{0}\right)\exp\left(2\pi ikx\right)+\sum_{k=-\infty}^{\infty}b_{k}\left(x_{0}\right)\exp\left(2\pi ikx\right),\label{alpha_b}
\end{equation}
where

\begin{equation}
b_{k}\left(x_{0}\right)=\sum_{m=-\infty}^{\infty}a_{\left(k-m\right)m}\left(x_{0}\right)\exp\left(-2\pi imx_{0}\right).\label{b_k_x_0}
\end{equation}
Suppose that $\lim_{x\rightarrow\infty}s\left(x\right)=0$. Then it
follows from (\ref{alpha_b}) that
$\lim_{x_{0}\rightarrow\infty}\alpha_{k}\left(x_{0}\right)=0$ and
$\lim_{x_{0}\rightarrow\infty}b_{k}\left(x_{0}\right)=0$ for all
$k$. The last equality and (\ref{b_k_x_0}) imply that
$\lim_{x_{0}\rightarrow\infty}a_{\left(k-m\right)k}\left(x_{0}\right)=0$
for all $k$ and $m$. Then it follows from (\ref{c_n}) that
$\lim_{x\rightarrow\infty}c_{n}\left(x\right)=0$ for all $n$,
which contradicts the conditions of the lemma. The contradiction
obtained proves Lemma \ref{lemma_1}.

Let us proceed to the proof of Theorem \ref{th_2}. Define
functions $a\left(x\right)$ and $b\left(x\right)$ from the class
$C^{\infty}\left(\mathbb{R}\right)$ so that the conditions
\[
a\left(n\right)=a_{n},\;b\left(n\right)=b_{n}\;\mathrm{for}\;n\geq0,
\]
\[
a\left(x\right)=a^{-x},\;b\left(x\right)=b^{-x}\;\mathrm{for}\;x\leq-1,
\]
are satisfied and the functions $a\left(x\right)a^{x}$ and
$b\left(x\right)b^{x}$ either are bounded for $x>0$ or tend to $1$
under condition (\ref{a_n_b_n}). Introduce the functions
\[
\varepsilon\left(x\right)=\dfrac{a\left(x\right)-a^{-x}}{a^{-x}},\;\delta\left(x\right)=\dfrac{b\left(x\right)-b^{-x}}{b^{-x}},
\]
which are bounded for all $x$, and, moreover, are infinitesimal as
$x\rightarrow+\infty$ under conditions (\ref{a_n_b_n}). Consider
the series
\begin{equation}
Q\left(t\right)=\mathop{\sum}\limits _{m=-\infty}^{\infty}a\left(m\right)e^{-b\left(m\right)t}=S\left(t\right)+R\left(t\right),\label{Q}
\end{equation}
where
\[
R\left(t\right)=\mathop{\sum}\limits _{i=1}^{\infty}a^{i}e^{-b^{i}t}.
\]
Define the functions $q\left(x\right)$ and $\bar{q}\left(x\right)$
as
\begin{equation}
q\left(x\right)=a\left(x\right)e^{-b\left(x\right)t}\in L^{1}\left(\mathbb{R}\right),\label{q_a_b}
\end{equation}
\begin{equation}
\bar{q}\left(x\right)=\sum_{n=-\infty}^{\infty}q\left(x+n\right).\label{q_}
\end{equation}
It is obvious that the function $\bar{q}\left(x\right)$ is
periodic with period $1$, integrable on any period, and satisfies
the Dirichlet condition. Then the following function is defined:
\[
\tilde{q}\left(m\right)\equiv\intop_{0}^{1}\bar{q}\left(x\right)\exp\left(-2\pi imx\right)dx
\]
\[
=\intop_{0}^{1}\sum_{n=-\infty}^{\infty}q\left(x+n\right)\exp\left(-2\pi imx\right)dx=\intop_{-\infty}^{\infty}q\left(x\right)\exp\left(-2\pi imx\right)dx,
\]
where
\begin{equation}
\bar{q}\left(x\right)=\sum_{m=-\infty}^{\infty}\tilde{q}\left(m\right)\exp\left(2\pi imx\right),\label{q(x)}
\end{equation}
and series (\ref{q(x)}) converges uniformly in $x$. It follows
from (\ref{q(x)}) that

\begin{equation}
\sum_{n=-\infty}^{\infty}q\left(x+n\right)=\sum_{m=-\infty}^{\infty}\tilde{q}\left(m\right)\exp\left(2\pi imx\right).\label{q_x+n}
\end{equation}
Setting $x=0$ in (\ref{q_x+n}), we obtain

\begin{equation}
\sum_{n=-\infty}^{\infty}q\left(n\right)=\sum_{m=-\infty}^{\infty}\intop_{-\infty}^{\infty}q\left(x\right)\exp\left(-2\pi imx\right)dx.\label{Pois}
\end{equation}
The substitution of (\ref{q_a_b}) into (\ref{Pois}) yields
\begin{equation}
Q\left(t\right)=\sum_{n=-\infty}^{\infty}a\left(n\right)e^{-b\left(n\right)t}=\sum_{m=-\infty}^{\infty}\intop_{-\infty}^{\infty}a\left(x\right)e^{-b\left(x\right)t}\exp\left(-2\pi
imx\right)dx.\label{Q(t)}
\end{equation}
Changing the variable $x\rightarrow y=-x\log b+\log t$ in
(\ref{Q(t)}), we obtain
\begin{equation}
Q\left(t\right)=\dfrac{1}{\log b}t^{-\tfrac{\log a}{\log b}}\mathop{\sum}\limits _{k=-\infty}^{+\infty}t^{\tfrac{2\pi ik}{\log b}}Q_{k}\left(t\right),\label{Q_t}
\end{equation}

\noindent where
\[
Q_{k}\left(t\right)=\mathop{\smallint}\limits _{-\infty}^{\infty}dy\exp\left(-\dfrac{2\pi ik}{\log b}y\right)
\]

\noindent
\begin{equation}
\times\left(1+\varepsilon\left(-\dfrac{y}{\log b}+\dfrac{\log t}{\log b}\right)\right)\exp\left(-e^{y}\delta\left(-\dfrac{y}{\log b}+\dfrac{\log t}{\log b}\right)\right)\exp\left(\dfrac{\log a}{\log b}y-e^{y}\right).\label{Q_k_y}
\end{equation}
Since

\[
1+\delta\left(-\dfrac{y}{\log b}+\dfrac{\log t}{\log b}\right)>0,\;1+\varepsilon\left(-\dfrac{y}{\log b}+\dfrac{\log t}{\log b}\right)>0,
\]
the integral (\ref{Q_k_y}) converges absolutely for any
$t>T\in\mathbb{R}_{+}$ and is the Fourier integral of the function
in the second row of (\ref{Q_k_y}). This function is a function
from the Schwartz space, i.e., a function of the class
$C^{\infty}$, which decreases together with all its derivatives as
$\left|y\right|\rightarrow\infty$ faster than
$\left|y\right|^{-m}$ for any $m>0$. Thus, $Q_{k}\left(t\right)$
for any $t>T\in\mathbb{R}_{+}$ is also a function from the
Schwartz class with respect to $k$ and decreases together with all
its derivatives as $\left|k\right|\rightarrow\infty$ faster than
$\left|k\right|^{-m}$ for any $m>0$. This means that the series
\begin{equation}
\log bt^{\tfrac{\log a}{\log b}}Q\left(t\right)=\mathop{\sum}\limits _{k=-\infty}^{+\infty}t^{\tfrac{2\pi ik}{\log b}}Q_{k}\left(t\right)\label{series_unif_t}
\end{equation}
converges absolutely for any $t>T\in\mathbb{R}_{+}$, i.e., the
series
\[
\mathop{\sum}\limits
_{k=-\infty}^{+\infty}\left|Q_{k}\left(t\right)\right|
\]
converges. Since $Q_{k}\left(t\right)$ is bounded with respect to
$t$, it follows that
\[
\mathop{\sum}\limits _{k=-\infty}^{+\infty}\left|Q_{k}\left(t\right)\right|\leq\mathop{\sum}\limits _{k=-\infty}^{+\infty}\sup_{t>T}\left|Q_{k}\left(t\right)\right|,
\]
which implies the boundedness and hence the uniform convergence in
$t>T$ of the series $\mathop{\sum}\limits
_{k=-\infty}^{+\infty}t^{\tfrac{2\pi ik}{\log
b}}Q_{k}\left(t\right)$. It follows from representation
(\ref{Q_k_y}) that the functions $Q_{k}\left(t\right)$ are not
infinitesimal as $t\rightarrow\infty$. Then, according to Lemma
\ref{lemma_1}, we have the representation
\begin{equation}
Q\left(t\right)=t^{-\tfrac{\log a}{\log b}}f\left(t\right),\label{Q_restr}
\end{equation}
where $f\left(t\right)$ is a bounded function that is not
infinitesimal as $t\rightarrow\infty$. Next, it is easy to show
that the series $t^{\tfrac{\log a}{\log
b}}R\left(t\right)=\mathop{\sum}\limits
_{i=1}^{\infty}a^{i}\exp\left(-b^{i}t+\log t\dfrac{\log a}{\log
b}\right)$ converges uniformly in $t$ and has zero limit as
$t\rightarrow\infty$, whence it follows that
\begin{equation}
Q\left(t\right)\sim S\left(t\right)\label{QS}
\end{equation}
as $t\rightarrow\infty$. Formula (\ref{QS}) and representation
(\ref{Q_restr}) imply the first assertion of Theorem \ref{th_2}.

Let us prove the second assertion of Theorem \ref{th_2}. Consider
the series

\begin{equation}
P\left(t\right)=\mathop{\sum}\limits _{m=-\infty}^{\infty}a^{-i}e^{-b^{-i}t},\label{P}
\end{equation}
which is a particular case of series (\ref{Q}). Then by
(\ref{Q_t})--(\ref{Q_k_y}) we have

\begin{equation}
P\left(t\right)=\dfrac{1}{\log b}t^{-\tfrac{\log a}{\log b}}\mathop{\sum}\limits _{k=-\infty}^{+\infty}t^{\tfrac{2\pi ik}{\log b}}P_{k}\left(t\right),\label{P_t}
\end{equation}
where

\noindent
\[
P_{k}\left(t\right)=\mathop{\smallint}\limits _{-\infty}^{\infty}dy\exp\left(\dfrac{\log a}{\log b}y-e^{y}\right)\exp\left(-\dfrac{2\pi ik}{\log b}y\right)
\]

\noindent
\begin{equation}
=\mathop{\smallint}\limits _{0}^{\infty}dx\exp\left(-x\right)x^{\tfrac{\log a-2\pi ik}{\log b}-1}=\Gamma\left(\dfrac{\log a}{\log b}-\dfrac{2\pi im}{\log b}\right)\label{P_k}
\end{equation}
and the series
\[
\mathop{\sum}\limits _{k=-\infty}^{+\infty}t^{\tfrac{2\pi ik}{\log b}}P_{k}\left(t\right)=\mathop{\sum}\limits _{k=-\infty}^{+\infty}t^{\tfrac{2\pi ik}{\log b}}\Gamma\left(\dfrac{\log a}{\log b}-\dfrac{2\pi im}{\log b}\right),
\]
just as the series (\ref{series_unif_t}), converges uniformly in
$t>T$. Consider the limit of the difference
\[
\lim_{t\rightarrow\infty}\left(\log bt^{\tfrac{\log a}{\log b}}Q\left(t\right)-\log bt^{\tfrac{\log a}{\log b}}P\left(t\right)\right)
\]
\[
=\lim_{t\rightarrow\infty}\mathop{\sum}\limits _{k=-\infty}^{+\infty}t^{\tfrac{2\pi ik}{\log b}}\mathop{\smallint}\limits _{0}^{\infty}dy\exp\left(-e^{y}-\dfrac{\log a}{\log b}y-\dfrac{2\pi ik}{\log b}y\right)
\]
\begin{equation}
\times\left(\left(1+\varepsilon\left(-\dfrac{y}{\log b}+\dfrac{\log t}{\log b}\right)\right)\exp\left(-e^{y}\delta\left(-\dfrac{y}{\log b}+\dfrac{\log t}{\log b}\right)\right)-1\right).\label{lim_diff}
\end{equation}

\noindent In view of the uniform convergence in $t$, we can take
the limit under the summation sign. Since

\noindent
\[
\lim_{t\rightarrow\infty}t^{\tfrac{2\pi ik}{\log
b}}\left(\left(1+\varepsilon\left(-\dfrac{y}{\log b}+\dfrac{\log
t}{\log
b}\right)\right)\exp\left(-e^{y}\delta\left(-\dfrac{y}{\log
b}+\dfrac{\log t}{\log b}\right)\right)-1\right)=0
\]
for any $y$, the limit (\ref{lim_diff}) is zero; hence,
\[
\lim_{t\rightarrow\infty}\left(\dfrac{Q\left(t\right)}{P\left(t\right)}-1\right)=\lim_{t\rightarrow\infty}\dfrac{\log bt^{\tfrac{\log a}{\log b}}Q\left(t\right)-\log bt^{\tfrac{\log a}{\log b}}P\left(t\right)}{\log bt^{\tfrac{\log a}{\log b}}P\left(t\right)}=0.
\]
It follows from the last equality that $Q\left(t\right)\sim
P\left(t\right)$; hence, taking into account (\ref{QS}), we obtain
\[
S\left(t\right)\sim P\left(t\right),
\]
which, with regard to (\ref{P_t})--(\ref{P_k}), proves the second
assertion of Theorem \ref{th_2}.

\section*{Appendix 3. Proof of Theorem \ref{th_4}}

\label{App_3}

In this appendix we present the proof of Theorem \ref{th_4}. From
equation (\ref{eq_pol}) we have
\[
1+kJ\left(-\nu_{j}\right)=0,\;,j=1,2,\ldots,
\]
which is equivalent to

\begin{equation}
\sum_{i=1}^{\infty}N_{i-1}^{-1}\left(1-\dfrac{N_{i-1}}{N_{i}}\right)\dfrac{1}{-\nu_{j}+\lambda_{i}}=-\dfrac{1}{k}.\label{eq_nu_k}
\end{equation}
Separating the $k$th term in the sum on the left-hand side of
equation (\ref{eq_nu_k}), we can rewrite (\ref{eq_nu_k}) as
\[
\dfrac{1}{\Delta_{j}}=\sum_{i=1}^{j-1}\dfrac{N_{i-1}^{-1}\left(1-\dfrac{N_{i-1}}{N_{i}}\right)}{N_{j-1}^{-1}\left(1-\dfrac{N_{j-1}}{N_{j}}\right)}\dfrac{1}{\dfrac{\lambda_{i}}{\lambda_{j}}-1-\Delta_{j}}-
\]
\begin{equation}
-\sum_{i=j+1}^{\infty}\dfrac{N_{i-1}^{-1}\left(1-\dfrac{N_{i-1}}{N_{i}}\right)}{N_{j-1}^{-1}\left(1-\dfrac{N_{j-1}}{N_{j}}\right)}\dfrac{1}{1-\dfrac{\lambda_{i}}{\lambda_{j}}+\Delta_{j}}+\dfrac{1}{k}N_{j-1}\left(1-\dfrac{N_{j-1}}{N_{j}}\right)^{-1}\lambda_{j}.\label{Delta_k}
\end{equation}

Let us prove the first assertion of Theorem \ref{th_4}. Suppose
that $\alpha>\dfrac{\theta}{\xi}$ and conditions
(\ref{Cond_N_restr}) and (\ref{Cond_d_restr}) are satisfied. Let
us show that the sequence $\Delta_{j}$ is bounded below by a
positive number. To this end, we consider the sequence
$\dfrac{1}{\Delta_{j}}$. Since the expressions under the summation
sign in (\ref{Delta_k}) are positive, we can make the following
estimate:

\[
\dfrac{1}{\Delta_{j}}<\sum_{i=1}^{j-1}\dfrac{N_{i-1}^{-1}\left(1-\dfrac{N_{i-1}}{N_{i}}\right)}{N_{j-1}^{-1}\left(1-\dfrac{N_{j-1}}{N_{j}}\right)}\dfrac{1}{\dfrac{\lambda_{i}}{\lambda_{j}}-1-\Delta_{j}}+\dfrac{1}{k}N_{j-1}\left(1-\dfrac{N_{j-1}}{N_{j}}\right)^{-1}\lambda_{j}<
\]
\begin{equation}
<\sum_{i=1}^{j-1}\dfrac{N_{j-1}\left(1-\dfrac{N_{j-1}}{N_{j}}\right)^{-1}\lambda_{j}}{N_{i-1}\left(1-\dfrac{N_{i-1}}{N_{i}}\right)^{-1}\lambda_{i}}+\dfrac{1}{k}N_{j-1}\left(1-\dfrac{N_{j-1}}{N_{j}}\right)^{-1}\lambda_{j}.\label{Ots}
\end{equation}
Representation (\ref{expr_lambda}), combined with the relation
$e^{-\theta}\dfrac{1+\varepsilon_{i}}{1+\varepsilon_{i+1}}<1$
obtained from (\ref{D_i_N_i}), implies that
$\lambda_{i}=e^{-\alpha\xi i}\sigma_{i},$ where $\sigma_{i}$ is a
positive bounded sequence. Then, using the second equation in
(\ref{D_i_N_i}), we obtain
\begin{equation}
N_{j-1}\left(1-\dfrac{N_{j-1}}{N_{j}}\right)^{-1}\lambda_{j}=Ce^{-\theta}\left(1+\varepsilon_{j}\right)\left(1-e^{-\theta}\dfrac{1+\varepsilon_{j-1}}{1+\varepsilon_{j}}\right)^{-1}e^{-\left(\alpha\xi-\theta\right)j}\sigma_{j}.\label{N_lambda_sigma}
\end{equation}
Taking into account (\ref{N_lambda_sigma}), we have the following
expression for the first term in (\ref{Ots}):
\[
\sum_{i=1}^{j-1}\dfrac{N_{j-1}\left(1-\dfrac{N_{j-1}}{N_{j}}\right)^{-1}\lambda_{j}}{N_{i-1}\left(1-\dfrac{N_{i-1}}{N_{i}}\right)^{-1}\lambda_{i}}
\]
\[
=\sum_{i=1}^{j-1}\dfrac{\left(1+\varepsilon_{j}\right)\left(1-e^{-\theta}\dfrac{1+\varepsilon_{j-1}}{1+\varepsilon_{j}}\right)^{-1}\sigma_{j}e^{-\left(\alpha\xi-\theta\right)j}}{\left(1+\varepsilon_{i}\right)\left(1-e^{-\theta}\dfrac{1+\varepsilon_{i-1}}{1+\varepsilon_{i}}\right)^{-1}\sigma_{i}e^{-\left(\alpha\xi-\theta\right)i}}
\]
\begin{equation}
\leq\dfrac{\sup_{i}\left(\left(1+\varepsilon_{i}\right)\left(1-e^{-\theta}\dfrac{1+\varepsilon_{i-1}}{1+\varepsilon_{i}}\right)^{-1}\sigma_{i}\right)}{\inf_{i}\left(\left(1+\varepsilon_{i}\right)\left(1-e^{-\theta}\dfrac{1+\varepsilon_{i-1}}{1+\varepsilon_{i}}\right)^{-1}\sigma_{i}\right)}\sum_{i=1}^{j-1}\dfrac{e^{-\left(\alpha\xi-\theta\right)j}}{e^{-\left(\alpha\xi-\theta\right)i}}.\label{sum_1}
\end{equation}
Since
\[
\sum_{i=1}^{j-1}\dfrac{e^{-\left(\alpha\xi-\theta\right)j}}{e^{-\left(\alpha\xi-\theta\right)i}}=e^{\alpha\xi-\theta}\dfrac{e^{\alpha\xi-\theta}-e^{-\left(\alpha\xi-\theta\right)j}}{e^{\left(\alpha\xi-\theta\right)}-1}<\dfrac{e^{2\left(\alpha\xi-\theta\right)}}{e^{\left(\alpha\xi-\theta\right)}-1},
\]
the sum (\ref{sum_1}) is bounded above by a positive number. For
the second term in (\ref{Ots}) we have
\[
\dfrac{1}{k}N_{j-1}\left(1-\dfrac{N_{j-1}}{N_{j}}\right)^{-1}\lambda_{j}
\]

\begin{equation}
\leq\dfrac{Ce^{-\theta}}{k}e^{-\left(\alpha\xi-\theta\right)}\sup_{i}\left(\left(1+\varepsilon_{i}\right)\left(1-e^{-\theta}\dfrac{1+\varepsilon_{i-1}}{1+\varepsilon_{i}}\right)^{-1}\sigma_{i}\right)>0.\label{N_lambda}
\end{equation}
Thus, we have shown that the sequence $\dfrac{1}{\Delta_{j}}$ is
bounded above by some positive number; hence, there exists a $c>0$
such that the following inequality holds for all $j>0$:
\begin{equation}
\Delta_{j}>c.\label{Del>0}
\end{equation}

Let us prove the first assertion of Theorem \ref{th_4}. Suppose
that conditions (\ref{Cond_N_restr})--(\ref{Cond_d_restr}) are
satisfied. The boundedness of the sequence $\Delta_{i}$ along with
(\ref{nu_lambda_delta}) and (\ref{expr_lambda}) imply that
$\dfrac{\nu_{j}}{e^{-\alpha\xi j}}$ is a bounded sequence. It
follows from (\ref{expr_lambda}) and (\ref{Cond_N_restr}) that the
sequences $\lambda_{j}e^{\left(\alpha\xi\right)j}$ and
$N_{j-1}\left(1-\dfrac{N_{j-1}}{N_{j}}\right)^{-1}\lambda_{j}e^{\left(\alpha\xi-\theta\right)j}$
are also bounded. Next, it follows from (\ref{b_k}) that
\[
b_{j}=\dfrac{k^{-2}}{\left|J^{\prime}\left(-\nu_{j}\right)\right|}=-\dfrac{k^{-2}}{\sum_{i=1}^{\infty}N_{i-1}^{-1}\left(1-\dfrac{N_{i-1}}{N_{i}}\right)\dfrac{1}{\left(\nu_{j}-\lambda_{i}\right)^{2}}}
\]
\[
=\dfrac{k^{-2}}{\sum_{i=1}^{\infty}N_{i-1}^{-1}\left(1-\dfrac{N_{i-1}}{N_{i}}\right)\dfrac{1}{\left(\lambda_{j}\Delta_{j}+\lambda_{j}-\lambda_{i}\right)^{2}}}.
\]
Therefore,
\[
b_{j}\sim\dfrac{k^{-2}\lambda_{j}^{2}\Delta_{j}^{2}}{N_{j-1}^{-1}\left(1-\dfrac{N_{j-1}}{N_{j}}\right)},
\]
\begin{equation}
\dfrac{b_{j}}{\nu_{j}}=\dfrac{b_{j}}{\lambda_{j}\left(1+\Delta_{j}\right)}\sim\dfrac{k^{-2}\lambda_{j}\Delta_{j}^{2}}{N_{j-1}^{-1}\left(1-\dfrac{N_{j-1}}{N_{j}}\right)\left(1+\Delta_{j}\right)}\label{b_del_nu}
\end{equation}
as $j\rightarrow\infty$. It follows from (\ref{b_del_nu}) that
under conditions (\ref{Cond_N_restr})--(\ref{Cond_d_restr}) the
sequence
$\dfrac{b_{j}}{\nu_{j}}e^{\left(\alpha\xi-\theta\right)j}$ is
bounded. Then, using the first assertion of Theorem \ref{th_2}, we
obtain the first assertion of Theorem \ref{th_4}.

Now, let us prove the second assertion of Theorem \ref{th_4}. To
this end, we first prove that under conditions
(\ref{Cond_N})--(\ref{Cond_d}) there exists a finite nonzero limit
$\lim_{j\rightarrow\infty}\Delta_{j}=\Delta.$ In view of
(\ref{Del>0}) and (\ref{Delta_restr}), the sequence
$\dfrac{1}{\Delta_{j}}$, $j>0$, is bounded above and below by two
positive numbers. Suppose that the limit
$\lim_{j\rightarrow\infty}\Delta_{j}$ does not exist. Then the
limit of the sequence $\dfrac{1}{\Delta_{j}}$ as
$j\rightarrow\infty$ does not exist either. In view of the
statement logically opposite to the Cauchy criterion, this means
that there exists a number $\varepsilon>0$ such that for any
$J\in\mathbb{Z}_{+}$ there exist $j,\:n\in\mathbb{Z}_{+}$, $j>J$,
such that
$\left|\dfrac{1}{\Delta_{j}}-\dfrac{1}{\Delta_{j+n}}\right|>\varepsilon$
(or $\left|\Delta_{j+n}-\Delta_{j}\right|>\varepsilon
c^{2}=\varepsilon^{\prime}$). Let us rewrite (\ref{Delta_k}) as

\[
\dfrac{1}{\Delta_{j}}=\sum_{i=1}^{j-1}\dfrac{N_{j-i-1}^{-1}\left(1-\dfrac{N_{j-i-1}}{N_{j-i}}\right)}{N_{j-1}^{-1}\left(1-\dfrac{N_{j-1}}{N_{j}}\right)}\dfrac{1}{\dfrac{\lambda_{j-i}}{\lambda_{j}}-1-\Delta_{j}}-
\]

\[
-\sum_{i=1}^{\infty}\dfrac{N_{i+j-1}^{-1}\left(1-\dfrac{N_{i+j-1}}{N_{i+j}}\right)}{N_{j-1}^{-1}\left(1-\dfrac{N_{j-1}}{N_{j}}\right)}\dfrac{1}{1-\dfrac{\lambda_{i+j}}{\lambda_{j}}+\Delta_{j}}+\dfrac{1}{k}N_{j-1}\left(1-\dfrac{N_{j-1}}{N_{j}}\right)^{-1}\lambda_{j}.
\]
Consider the difference
$\dfrac{1}{\Delta_{j}}-\dfrac{1}{\Delta_{j+n}}$. First, taking
into account (\ref{condition_lam}) and (\ref{condition_N}), we
write

\[
N_{j}^{-1}\left(1-\dfrac{N_{j}}{N_{j+1}}\right)=Ce^{-\theta j}\left(1-e^{-\theta}\right)\left(1+\omega_{j}\right),
\]

\[
\lambda_{j}=\eta e^{-\alpha D}e^{-\alpha\xi j}\left(1+\chi_{j}\right),
\]
where $\omega_{j}$ and $\chi_{j}$ are infinitesimal sequences:
$\lim_{i\rightarrow\infty}\omega_{j}=\lim_{i\rightarrow\infty}\chi_{j}=0$.
Then after some transformations we have:

\[
\dfrac{1}{\Delta_{j}}-\dfrac{1}{\Delta_{j+n}}
\]
\[
=-\sum_{i=j}^{j+n-1}\dfrac{1+\omega_{j+n-i-1}}{1+\omega_{j+n-1}}\dfrac{e^{\theta i}}{\dfrac{\left(1+\chi_{j+n-i}\right)}{\left(1+\chi_{i}\right)}e^{\alpha\xi i}-1-\Delta_{j+n}}
\]
\[
+\dfrac{1}{k}C^{-1}e^{-\theta}\left(1-e^{-\theta}\right)^{-1}\eta e^{-\alpha D}\left(1+\omega_{j-1}\right)^{-1}\left(1+\chi_{j}\right)e^{-\left(\alpha\xi-\theta\right)j}
\]
\[
-\dfrac{1}{k}C^{-1}e^{-\theta}\left(1-e^{-\theta}\right)^{-1}\eta e^{-\alpha D}\left(1+\omega_{j+n-1}\right)^{-1}\left(1+\chi_{j+n}\right)e^{-\left(\alpha\xi-\theta\right)\left(j+n\right)}
\]

\[
+\sum_{i=1}^{j-1}\left(\dfrac{1+\omega_{j-i-1}}{1+\omega_{j-1}}\dfrac{e^{\theta i}}{\dfrac{\left(1+\chi_{j-i}\right)}{\left(1+\chi_{i}\right)}e^{\alpha\xi i}-1-\Delta_{j}}-\dfrac{1+\omega_{j+n-i-1}}{1+\omega_{j+n-1}}\dfrac{e^{\theta i}}{\dfrac{\left(1+\chi_{j+n-i}\right)}{\left(1+\chi_{i}\right)}e^{\alpha\xi i}-1-\Delta_{j+n}}\right)
\]
\begin{equation}
-\sum_{i=1}^{\infty}\left(\dfrac{1+\omega_{j-i-1}}{1+\omega_{j-1}}\dfrac{e^{\theta
i}}{1-\dfrac{1+\chi_{j-i}}{1+\chi_{i}}e^{\alpha\xi
i}+\Delta_{j}}-\dfrac{1+\omega_{j+n-i-1}}{1+\omega_{j+n-1}}\dfrac{e^{\theta
i}}{1-\dfrac{1+\chi_{j+n-i}}{1+\chi_{i}}e^{\alpha\xi
i}+\Delta_{j+n}}\right).\label{neq}
\end{equation}
For sufficiently large $j$, the terms in the second, third, and
fourth rows on the right-hand side of (\ref{neq}) can be made
arbitrarily small independently of $\Delta_{j}$ and
$\Delta_{j+n}$. Conversely, the terms in the fifth and sixth rows
of (\ref{neq}) depend on the relationship between $\Delta_{j}$ and
$\Delta_{j+n}$. When
$\left|\Delta_{j+n}-\Delta_{j}\right|>\varepsilon^{\prime}$, the
absolute values of these terms cannot be made arbitrarily small in
absolute value for indefinitely increasing $j$, and they are
bounded from below. Moreover, notice that, for sufficiently large
$J$, the sign of the terms in the fifth and sixth rows of
(\ref{neq}) can always be made opposite to the sign of the
difference $\dfrac{1}{\Delta_{j}}-\dfrac{1}{\Delta_{j+n}}$. This
means that, for sufficiently large $J$, there always exist
$j,\:n\in\mathbb{Z}_{+}$, $j>J$, such that the right- and
left-hand sides (\ref{neq}) have different signs. This
contradiction proves the existence of the finite positive limit
$\lim_{i\rightarrow\infty}\Delta_{i}=\Delta>0$.

It follows from (\ref{b_del_nu}) with regard to
(\ref{condition_lam}) and from the existence of the limit
$\lim_{i\rightarrow\infty}\Delta_{i}=\Delta$ that
\begin{equation}
\nu_{j}=\lambda_{j}\left(1+\Delta_{j}\right)\sim\eta e^{-\alpha
D}\left(1+\Delta\right)e^{-\alpha\xi j}\label{nu}
\end{equation}
as $j\rightarrow\infty$. Then from (\ref{b_del_nu}) we obtain
\begin{equation}
\dfrac{b_{j}}{\nu_{j}}\sim\dfrac{k^{-2}A\Delta^{2}}{\left(1+\Delta\right)}e^{-\left(\alpha\xi-\theta\right)j}.\label{b_nu}
\end{equation}
Applying the second assertion of Theorem \ref{th_2} and taking
into account (\ref{b_nu})--(\ref{nu}), we obtain the second
assertion of Theorem \ref{th_4}.

\end{document}